\documentclass[sigconf,nonacm]{acmart}
\AtBeginDocument{%
  \providecommand\BibTeX{{%
    \normalfont B\kern-0.5em{\scshape i\kern-0.25em b}\kern-0.8em\TeX}}}

\usepackage{tcolorbox}
\usepackage[frozencache=true,cachedir=minted-cache]{minted} 
\newtcolorbox[list inside=mybox,auto counter]{codexexample}{colbacktitle=pink,coltitle=black, title={Codex Example \thetcbcounter ~(input in black, output in red)},float, floatplacement=t!}

\begin{document}

\title{Using Large Language Models to Enhance Programming~Error~Messages}


\author[Leinonen]{Juho Leinonen}
\orcid{0000-0001-6829-9449}
\affiliation{
  \institution{Aalto University}
  \city{Espoo}
  \country{Finland}
}
\email{juho.2.leinonen@aalto.fi}

\author[Hellas]{Arto Hellas}
\orcid{0000-0001-6502-209X}
\affiliation{
  \institution{Aalto University}
  \city{Espoo}
  \country{Finland}
}
\email{arto.hellas@aalto.fi}

\author[Sarsa]{Sami Sarsa}
\orcid{0000-0002-7277-9282}
\affiliation{
  \institution{Aalto University}
  \city{Espoo}
  \country{Finland}
}
\email{sami.sarsa@aalto.fi}

\author[Reeves]{Brent Reeves}
\orcid{0000-0001-5781-1136}
\affiliation{
  \institution{Abilene Christian University}
  \city{Abilene, Texas}
  \country{USA}
}
\email{brent.reeves@acu.edu}

\author[Denny]{Paul Denny}
\orcid{0000-0002-5150-9806}
\affiliation{
  \institution{The University of Auckland}
  \city{Auckland}
  \country{New Zealand}
}
\email{paul@cs.auckland.ac.nz}

\author[Prather]{James Prather}
\orcid{0000-0003-2807-6042}
\affiliation{
  \institution{Abilene Christian University}
  \city{Abilene, Texas}
  \country{USA}
}
\email{james.prather@acu.edu}

\author[Becker]{Brett A. Becker}
\orcid{0000-0003-1446-647X}
\affiliation{
  \institution{University College Dublin}
  \city{Dublin}
  \country{Ireland}
}
\email{brett.becker@ucd.ie}


\begin{abstract}
A key part of learning to program is learning to understand programming error messages. They can be hard to interpret and identifying the cause of errors can be time-consuming. One factor in this challenge is that the messages are typically intended for an audience that already knows how to program, or even for programming environments that then use the information to highlight areas in code. Researchers have been working on making these errors more novice friendly since the 1960s, however progress has been slow. The present work contributes to this stream of research by using large language models to enhance programming error messages with explanations of the errors and suggestions on how to fix the error. Large language models can be used to create useful and novice-friendly enhancements to programming error messages that sometimes surpass the original programming error messages in interpretability and actionability. These results provide further evidence of the benefits of large language models for computing educators, highlighting their use in areas known to be challenging for students. We further discuss the benefits and downsides of large language models and highlight future streams of research for enhancing programming error messages.  
\end{abstract}




\maketitle

\section{Introduction}
\label{sec:introduction}

Programming Error Messages (PEMs) can be notoriously difficult to decipher, especially for novices ~\cite{prather2017novices}, possibly to the extent that they contribute ot the perception that programming is overly challenging~\cite{becker2021what}. Eye tracking studies reveal that novices read error messages and spend a substantial amount of programming time trying to understand them ~\cite{barik2017do}. Instructors report that they spend a considerable amount of time helping novices with these often cryptic messages ~\cite{denny2011understanding, stefik2013empirical, pettit2017enhanced, prather2018metacognitive}. It is also known that error message presentation affects novice programming behaviour~\cite{karvelas2020effects}. For over six decades, researchers have attempted to improve these messages, and still there is a call for more work on the matter ~\cite{Becker2019wgpaper}. Some recent attempts have been made to put error messages into more natural language by focusing on an increase in readability ~\cite{denny2021designing, becker2021towards}. This has been shown to improve student understanding of error messages and the number of successful code corrections ~\cite{denny2020error}. While it is clear that increasing the readability of PEMs is helpful to novices, doing so at scale, and across languages, remains a challenge.

Very recent work on using large language models in computing education have already produced promising results. One study reported that Codex --  built on top of GPT-3 (see Section~\ref{LLM}) -- could solve introductory programming problems, and ranked Codex in the top quartile when compared to a cohort of actual students in a large introductory programming course~\cite{finnie2022robots}. Tools like Codex are able to generate new programming assignments~\cite{sarsa2022automatic} and code explanations~\cite{macneil2022generating} when provided examples. Such tools demonstrate the impressive interpretive power of very recent large language models that may have the potential to improve the readability of input text. In this paper, we investigate whether large language models can be utilized to parse non-compiling code and the programming errors generated from that code to output PEMs that are more readable than those generated by the compiler/interpreter.

\begin{itemize}
     \item [RQ1] How well can Codex explain different error messages?
     \item [RQ2] What is the quality of the code fix suggestions that Codex generates?
 \end{itemize}

\section{Background}
\label{sec:background}

\subsection{Programming Error Messages}

Programming error messages (PEMs) encompass syntax error messages, compiler error messages, and other diagnostic messages that are produced by compilers or interpreters indicating that the input code violates the specification of a language~\cite{becker2016effectiveCSEJ}. Researchers and instructors have reported PEMs to be a difficulty for students since at least 1965~\cite{rosen1965pufft}. More than fifty years later, PEMs are still a barrier to progress for those learning to program~\cite{Becker2019wgpaper}, and this has led to various efforts for improve their usability. 

One such avenue of work has involved intercepting messages between the compiler and the user and altering their wording or presentation.  One of the many known issues with error messages generated by compilers and interpreters is poor readability due to factors such as poor use of vocabulary, strange sentence structure, and the use of jargon~\cite{denny2021designing}.  Thus, a large body of work around so-called `enhanced compiler error messages' has emerged ~\cite{becker2016effectiveCSEJ}.  Different approaches to message wording have been reported by various authors, including Barik~\cite{barik2018error}, Becker~\cite{becker2018effects}, Denny~\cite{denny2014enhancing}, Kohn~\cite{kohn2019error}, Pettit~\cite{pettit2017enhanced}, Prather~\cite{prather2017novices}, and  Karkare~\cite{ahmed2018compilation}.  However, although some studies have shown positive effects of rewording messages for novices ~\cite{becker2016effective, denny2020error}, in general the evidence for the effectiveness of enhanced compiler error messages is not overwhelming.  One of the limitations of prior work in this area is the manual effort that is required to generate message rewordings and a lack of clear guidance for addressing core issues such as readability \cite{denny2021designing}.

Artificial intelligence and machine learning approaches have been used for finding and repairing errors in programs~\cite{ahmed2021synfix, gupta2017deepfix, gupta2019deep} but only very fundamental approaches have been applied to researching PEMs~\cite{becker2016categorizing}.  To our knowledge, no prior work has explored the use of large language models for improving PEMs.

\subsection{Large Language Models}
\label{LLM}

Large Language Models (LLMs), particularly pre-trained transformer models, have rapidly become the core technologies of natural language processing~\cite{hang2022language}. One such model is OpenAI GPT-3 (third-generation Generative Pre-trained Transformer)~\cite{brown2020language}. GPT-3 can translate between natural languages, compose poetry in the style of human poets, generate convincing essays, and more. GPT-3 also powers several other tools such as OpenAI Codex which is essentially a GPT-3 model that has also been fine-tuned with more than 50 million repositories representing the majority of Python code available on GitHub totalling 159\,GB of source code~\cite{chen2021evaluating}. Codex is available via the OpenAI API (\href{https://beta.openai.com/}{beta.openai.com}) and also powers tools such as GitHub Copilot (\href{https://copilot.github.com/}{copilot.github.com}).

Given the recent emergence of these models, little is yet known about the impact they are likely to have on the computing education landscape.  In this context, the few evaluations conducted to date have focused on the accuracy of such models for solving typical introductory programming problems and on the potential for the models to generate learning resources.  Early work by Finnie-Ansley et al. assessed the accuracy of Codex by presenting it with typical CS1-type problems, and comparing its performance against that of students.  They found that it outperformed most students, and was capable of generating a variety of correct solutions to any given problem \cite{finnie2022robots}.  Sarsa et al. investigated the content generation capabilities of Codex, by providing input examples as  prompts and using it to generate novel programming problems and code explanations \cite{sarsa2022automatic}.  They found that most of the problems generated by Codex were novel and sensible, and that the generated code explanations were generally correct and thorough. 

Given their capability for generating output of human-like quality from contextual inputs, such as code explanations from code, there is potential in applying large language models to the problem of enhancing PEMs.

\section{Methodology}
\label{sec:methodology}

\subsection{Error Messages and Programs}

For the present study, we collected Python error messages that had been reported as the most unreadable in~\cite{denny2021designing} and~\cite{becker2021towards}. These error messages were as follows:

\begin{enumerate}
    \item can't assign to function call
    \item invalid token
    \item illegal target for annotation
    \item unindent does no match any outer indentation level
    \item positional argument follows keyword argument
    \item unexpected EOF while parsing
    \item EOL while scanning string literal
    \item EOF while scanning triple-quoted string literal
    \item (unicodeerror) `unicodeescape' codec can't decode bytes
\end{enumerate}

To control whether the complexity of the program that results in a given error message affects the ability of large language models to create useful explanations of the message, we constructed three example programs that generated each error message. The first program was very simple, often only a few lines long. The second incorporated the usage of strings and functions. The third included the use of libraries (e.g., the PyGame game library, pandas, scikit-learn) and was more complex. To create the same error messages as in the works by~\cite{denny2021designing} and~\cite{becker2021towards}, we used Python version 3.6.

\subsection{Generating Programming Error Messages}

Programming error messages were generated using the Codex model that was most recent and performant at the time of analysis, which was the \texttt{code-davinci-002} -model. As the utility of large language models depends on the used prompts (see e.g.,~\cite{liu2021pre}), it is important to do ``prompt engineering'' where the performance of different types of prompts is evaluated~\cite{liu2021pre}. We evaluated a number of prompts to identify a version that seemed to provide useful explanations. We tried five different prompt messages:

\begin{itemize}
    \item[1.] Plain English explanation of why does running the above code cause an error and how to fix the problem
    \item[2.] Plain English explanation of why running the above code causes the above error in the output and instructions on how to fix the problem
    \item[3.] Explanation of why running the above code causes the above error and instructions on how to fix the problem
    \item[4.] Why does the code result in an error message? How can the code be fixed?
    \item[5.] Why does the above code cause the above error message in the output? How can the code be fixed?
\end{itemize}

We generated explanations with all five prompts and checked which version led to the fewest empty responses from Codex. The number of empty responses was 4, 6, 7, 16 and 27 out of 81 generated explanations respectively for the prompts 1 to 5 above. We chose the first for the analysis as it generated the fewest empty responses. The structure of the prompt given to the large language model can be seen in the Codex Examples provided later in this article.

For each error message (9 error messages) and each program leading to an error message (3 programs), we generated three code explanations, one with Codex temperature parameter set to 0, and two with temperature set to 0.7. We chose these values as 0 is the minimum for the parameter and leads to least randomness, i.e. most deterministic outputs. The value of 0.7 is the default value for the parameter and leads to more random (or ``creative'') responses, and is less deterministic, which is also why we generated two explanations for the value of 0.7. This led to a total of $9 \times 3 \times 3 = 81$ unique combinations of programming error message, program category, and temperature value, which we subsequently evaluated.

\subsection{Analysis}
\label{sec:analysis}

We qualitatively analyzed the LLM-produced PEMs. The evaluation was performed by two researchers, both of whom have experience from teaching introductory programming. For the evaluation, we considered the following aspects of the generated PEMs.

\begin{enumerate}
    \item \textit{Comprehensible}: was the generated content intelligible (i.e. proper English, not nonsensical) 
    \item \textit{Unnecessary content}: did the generated explanation contain unnecessary content (e.g., repeating content, comprehensible but irrelevant content) 
    \item \textit{Has explanation}: did the content produced by the LLM contain an explanation of the programming error message 
    \item \textit{Explanation correct}: did the content produced by the LLM contain a \emph{correct} explanation of the programming error message 
    \item \textit{Has fix}: did the generated explanation contain actions or steps that one should take to fix the error 
    \item \textit{Fix correct}: did the content produced by the LLM contain \emph{correct} actions or steps that one should take to fix the error 
    \item \textit{Improvement over the original}: did the explanation provide added value (from a novice programmer standpoint) when compared to the original programming error message 
\end{enumerate}

The researchers first had a brief discussion to ensure a shared understanding of the above aspects and jointly evaluated three examples. After the discussion and initial joint evaluation, they separately analyzed the full set of generated explanations. For each aspect, the researchers chose either ``yes'' or ``no''. For evaluation, the researchers also had access to the original error message as well as the program that produced the error message, and considered also these when evaluating the LLM generated explanations. To examine the validity of the approach, we calculated inter-rater reliability between the researchers using Cohen's kappa. The kappa value was 0.83 over all the analyzed aspects, indicating almost perfect agreement~\cite{landis1977measurement}.

To answer both of our research questions, we report the percentage of ``yes'' answers for the questions outlined above separately for each different programming error message and separately for each combination of program category and temperature value. The proportion of ``yes'' answers is calculated out of the full set of 162 data points: 2 raters, each with 81 distinct ratings for the unique combinations of programming error message (n = 9), program (n = 3), and Codex output (n = 3).

\section{Results}
\label{sec:results}

\begin{table*}[]
\centering
\resizebox{\textwidth}{!}{%
\begin{tabular}{@{}l|ccccc|cc@{}}
\toprule
& \multicolumn{5}{|c}{RQ1} & \multicolumn{2}{|c}{RQ2} \\
Error message                                           & \multicolumn{1}{|l}{Comprehensible} & \multicolumn{1}{l}{Unnecessary content} & \multicolumn{1}{l}{Has explanation} & \multicolumn{1}{l}{Explanation correct} & \multicolumn{1}{l}{Improvement} & \multicolumn{1}{|l}{Has fix} & \multicolumn{1}{l}{Fix correct} \\ \midrule
can’t assign to function call                           & 100\%                              & 17\%                                    & 94\%                                & 83\%                                    & 78\%                            & 72\%                        & 28\%                            \\
invalid token                                           & 100\%                              & 39\%                                    & 89\%                                & 50\%                                    & 78\%                            & 83\%                        & 44\%                            \\
illegal target for annotation                           & 67\%                               & 22\%                                    & 67\%                                & 33\%                                    & 33\%                            & 50\%                        & 28\%                            \\
unindent does not match any outer indentation level     & 100\%                              & 39\%                                    & 100\%                               & 56\%                                    & 56\%                            & 67\%                        & 28\%                            \\
positional argument follows keyword argument            & 89\%                               & 22\%                                    & 89\%                                & 61\%                                    & 56\%                            & 78\%                        & 39\%                            \\
unexpected EOF while parsing                            & 67\%                               & 11\%                                    & 67\%                                & 11\%                                    & 22\%                            & 44\%                        & 22\%                            \\
EOL while scanning string literal                       & 89\%                               & 28\%                                    & 89\%                                & 22\%                                    & 50\%                            & 67\%                        & 17\%                            \\
EOF while scanning triple-quoted string literal         & 89\%                               & 56\%                                    & 78\%                                & 44\%                                    & 44\%                            & 89\%                        & 33\%                            \\
(unicodeerror) `unicodeescape' codec can't decode bytes & 89\%                               & 56\%                                    & 83\%                                & 72\%                                    & 67\%                            & 78\%                        & 56\%                            \\
\midrule
Average over all error messages                                                   & 88\%                               & 32\%                                    & 84\%                                & 48\%                                    & 54\%                            & 70\%                        & 33\%                            \\ \bottomrule
\end{tabular}%
}
\caption{Error message analysis for each research question. The cells show the percentage of ``yes'' answers out of all (``yes'' and ``no'') answers for the analysis.}
\label{tab:results:pems}
\end{table*}

\begin{table*}[]
\centering
\resizebox{\textwidth}{!}{%
\begin{tabular}{@{}lc|ccccc|cc@{}}
\toprule
& & \multicolumn{5}{|c}{RQ1} & \multicolumn{2}{|c}{RQ2} \\
Program category      & Temperature & \multicolumn{1}{l}{Comprehensible} & \multicolumn{1}{l}{Unnecessary content} & \multicolumn{1}{l}{Has explanation} & \multicolumn{1}{l}{Explanation correct} & \multicolumn{1}{l}{Improvement} & \multicolumn{1}{|l}{Has fix} & \multicolumn{1}{l}{Fix correct} \\ \midrule
Simple                & 0.0         & 100\%                              & 6\%                                     & 100\%                               & 67\%                                    & 72\%                            & 78\%                        & 44\%                            \\
Function with strings & 0.0         & 100\%                              & 22\%                                    & 100\%                               & 56\%                                    & 72\%                            & 78\%                        & 33\%                            \\
Library               & 0.0         & 100\%                              & 28\%                                    & 100\%                               & 78\%                                    & 78\%                            & 72\%                        & 44\%                            \\
\midrule
Simple                & 0.7         & 83\%                               & 31\%                                    & 78\%                                & 47\%                                    & 42\%                            & 64\%                        & 31\%                            \\
Function with strings & 0.7         & 89\%                               & 42\%                                    & 86\%                                & 36\%                                    & 39\%                            & 75\%                        & 25\%                            \\
Library               & 0.7         & 72\%                               & 44\%                                    & 64\%                                & 33\%                                    & 50\%                            & 61\%                        & 31\%                            \\ \bottomrule
\end{tabular}%
}
\caption{Effect of temperature and program category on Codex performance in the task.}
\label{tab:results:tempcomp}
\end{table*}

Table~\ref{tab:results:pems} shows the results of the analysis separately for each error message. Each cell of the table presents the percentage of ``yes'' answers to the evaluation question (see Section~\ref{sec:analysis} for the questions) for each of the nine error messages. The cells in the bottom row of the table show the percentage of ``yes'' answers across all error messages for the evaluation question indicated by the column.

In general, most error message explanations created by Codex were comprehensible (percentage of ``yes'' ranging from 67\% to 100\%). A few of the created explanations contained unnecessary content such as repeated sentences, extra question marks, etc -- the percentage ranging from 11\% for ``unexpected EOF while parsing'' to 56\% for ``EOF while scanning triple-quoted string literal'' and ``(unicodeerror) `unicodeescape' codec can't decode bytes''.

In most cases, Codex successfully created an explanation of the error message (67\% to 100\% of the time depending on error message), although there were considerable differences between error messages on whether the explanation was correct. The range of correct explanations ranged from 11\% for ``unexpected EOF while parsing'' to 83\% for ``can't assign to function call''.

Regarding Codex's ability to create actionable fixes based on the faulty source code and the programming error message, we found that in the majority of cases, Codex provided a fix in the generated explanation (44\% to 89\% of cases). However, the fix was correct only 33\% of the time, ranging from 17\% of the time for ``EOL while scanning string literal'' to 56\% for ``(unicodeerror) `unicodeescape' codec can't decode bytes''.

Altogether, the evaluators considered that the Codex-created content, i.e. the explanation of the error message and the proposed fix, were an improvement over the original error message in slightly over half of the cases (54\%). There were some differences between different error messages: the content was an improvement only 22\% of the time for the ``unexpected EOF while parsing'' error message, while it was considered an improvement in 78\% of the cases for ``can't assign to function call'' and ``invalid token''.

Table~\ref{tab:results:tempcomp} shows the results of the analysis separately for different combinations of program category and temperature value. From the table, it is evident that for the task of explaining PEMs and creating suggestions for fixes to the source code that produced those errors, using a temperature value of 0 resulted in considerably better outputs, which holds for all three program categories. For example, the output was considered an improvement in over 70\% of the cases with a temperature value of 0, while only up to 50\% of the cases with a temperature value of 0.7.

Regarding differences between program categories, we can observe that Codex seems to perform slightly worse with the programs in the ``function with strings'' category. However, the variations between program categories are not as noticeable as the differences between different temperature values or different error messages.

\section{Discussion}
\label{sec:discussion}

\subsection{Are Error Message Explanations Useful?}

Our results suggest that using large language models to explain programming error messages (PEMs) is feasible and shows promise.  Overall, the explanation was considered an improvement over the original programming error message in over half of the cases. If we only consider the results from using temperature value 0, which were overall better, over 70\% of the Codex outputs were considered an improvement over the original programming error message.

The results are more sobering when it comes to using large language models to generate correct fixes.  Although 70\% included a fix, when a fix was included it was correct only under half of the time (47\%). While the fixes created with the temperature value of 0 were better on average compared to those created with a value of 0.7, they were still correct in only around half of the cases -- 42\%, 56\%, and 61\% of the cases for functions with strings, simple programs, and library related programs respectively.

We propose that the generated content could be useful to students if it were delivered so that it is clear that the content is AI-generated and might not be correct.   And even with this initial  exploratory setup, we found some outputs (two examples seen in Codex Example 1 and Codex Example 2) that we consider good enough to be shown to students without modifications.

% An example of a good explanation and fix
\begin{codexexample}
\begin{Verbatim}[fontsize=\small,breaklines,commandchars=\\\{\}]
""" Code
def check_password(password, input):
  return password == input
input = "hunter2"
if check_password("s3cr37", input):
""" Output
  File "main.py", line 5
    
                                       ^
SyntaxError: unexpected EOF while parsing
""" Plain English explanation of why does running the above code cause an error and how to fix the problem
\textcolor{red}{The error is caused because the block of code is expecting another line of code after the colon.}
\textcolor{red}{To fix the issue, I would add another line of code after the colon.}
\end{Verbatim}
\end{codexexample}

% Another example of a good explanation and fix
\begin{codexexample}
\begin{Verbatim}[fontsize=\small,breaklines,commandchars=\\\{\}]
""" Code
users_dir_path = "C:\textbackslash{}Users"
print("Users directory is", users_dir_path)
""" Output
  File "main.py", line 1
    users_dir_path = "C:\textbackslash{}Users"
                    ^
SyntaxError: (unicode error) `unicodeescape' codec can't decode bytes in position 2-3: truncated \textbackslash{}UXXXXXXXX escape
""" Plain English explanation of why does running the above code cause an error and how to fix the problem
\textcolor{red}{The error is caused by the backslash character in the string.}
\textcolor{red}{The backslash character is used to escape characters that would otherwise have a special meaning, such as newline, backslash itself, or the quote character.}
\textcolor{red}{Since the backslash is used for escaping, if you want to use it in your string, you need to escape it.}
\textcolor{red}{You can do that by using a double backslash:}
\end{Verbatim}
\end{codexexample}

\subsection{Common Pitfalls and Ways Around Them}

% Example of a confidently incorrect explanation and fix
\begin{codexexample}
\begin{Verbatim}[fontsize=\small,breaklines,commandchars=\\\{\}]
""" Code
import pygame

display = pygame.display.set_mode((640, 400))
pygame.display.set_caption("Awesome game!")

def main():
    while True:
""" Output
  File "main.py", line 8
    
                   ^
SyntaxError: unexpected EOF while parsing
""" Plain English explanation of why does running the above code cause an error and how to fix the problem
\textcolor{red}{# The above code is not indented enough to be able to be run by Python.}
\textcolor{red}{# Add more spaces or tabs to the start of each line of code to be able to run it.}
\end{Verbatim}
\end{codexexample}

% Example where both the explanation and fix are wrong
\begin{codexexample}
\begin{Verbatim}[fontsize=\small,breaklines,commandchars=\\\{\}]
""" Code
def check_password(password, input):
  If (input == "s3cr37"): print("You are in!")
  Else: print("Wrong password!")
input = "hunter2"
check_password("s3cr37", input)
""" Output
  File "main.py", line 2
    If (input == "s3cr37"): print("You are in!")
    ^
SyntaxError: illegal target for annotation
""" Plain English explanation of why does running the above code cause an error and how to fix the problem
\textcolor{red}{The error is caused by the fact that the code is not indented properly. The code should be indented by 4 spaces.}
\end{Verbatim}
\end{codexexample}

Two examples of outputs where both the explanation and suggested fix generated by Codex were incorrect are shown in Codex Examples 3 and 4. Comparing these incorrect outputs with the correct outputs in Examples 1 and 2, we observe that the messages seem similarly confident in their tone, which could potentially mislead students. In both of the examples where the output is incorrect, Codex suggests that the issue is related to indentation. As novices often struggle with indentation~\cite{liu2019static,kohn2019error}, these incorrect suggestions could exacerbate this by potentially misleading students and even introduce misconceptions related to correct indentation.

In general, we observed a few common pitfalls that Codex seemed to often struggle with: (1) source code clearly missing a part of the content (resulting in ``unexpected EOF while parsing'', see e.g., Codex Example 3), (2) incorrectly capitalized control statements (resulting in ``illegal target for annotation'', see e.g., Codex Example 4), and (3) missing quotation marks (resulting in either ``EOL while scanning string literal'' or ``EOF while scanning triple-quoted string literal'').

For the first case, Codex would often suggest to fix the indentation of the program, even though the problem was that the implementation was far from complete (as in Codex Example 3). Similar suggestions for fixing the indentation were observed for the second case as well, even though the problem is in the capitalization. This can be seen in Codex Example 4, where the issue is that the if-statement is capitalized, but the message claims the issue is with indentation. For the third case, Codex was often unable to correctly identify whether the quotation mark was missing from the beginning or the end of the string, and sometimes suggested that the issue is related to parentheses instead of missing quotation marks. Indeed, the program category ``function with strings'' had the lowest scores overall (see Table~\ref{tab:results:tempcomp}).

While it was relatively rare, we did observe some outputs that were not just incorrect, but even contradictory and confusing. In one case, Codex seems to have focused too much on the ``Plain English'' portion of the input and started generating irrelevant content related to ``looking for a plain English explanation''. To add to the confusion, the generated output actually does include a correct explanation of the problem -- ``You need to end your string with three single quotes at the end of your string to make it work.'', but the output also states that ``this is not a correct explanation''.

As there were common pitfalls and clear differences between explanation quality, we see one stream of future work in using a two-tiered approach for creating explanations. Codex could be relied upon in cases where it is known that it likely performs well, while in other cases other means could be exercised. One possibility is using LLMs to pre-generate explanations of common error messages that the instructor could validate (essentially, a ``human-in-the-loop'' approach). Another possibility would be the use of learnersourcing, where students could ask for help from their peers; classic approaches such as discussion forums would also work, although the response times would be lower when compared to the near-instantaneous feedback from Codex.

\subsection{Explanations and Context}

When considering the usefulness of Codex-generated explanations, they need to be interpreted and evaluated in context. First, the original error messages might be more useful for more experienced students who have learned to interpret them. The importance of context was present also in some of the disagreements of the two researchers who independently evaluated the error messages; for example, one of the researchers at times considered the error message as an improvement if it pointed the students to the correct direction, even if the explanation by itself would be faulty.

The utility of these explanations also depends on whether students understand the implications of the suggestions. Prior research into LLMs has shown that when they are used to facilitate the creation of source code, they may lead students down debugging rabbit holes~\cite{vaithilingam2022expectation} or even introduce security flaws~\cite{pearce2022asleep}. We also see the potential for other types of LLM problems. For example, what if the problem is not with the source code, but an issue with the user environment -- here, a student could ask for help to fix an issue, convincing the LLM that an issue exists, and going down a rabbit hole when looking for a solution~\cite{vaithilingam2022expectation}.

Despite the shortcomings, we see the potential of using LLMs as a scaffold when learning to program and when learning to interpret error messages. However, as with any instructional scaffolding, the scaffolding should be dismantled at some point~\cite{kalyuga2009expertise}, and students must eventually learn to understand the original error messages.

\subsection{Limitations}

There are limitations to our study, which we outline here. Firstly, we used Python 3.6 in the analysis similar to prior work~\cite{denny2021designing}. On one hand, this allowed us to focus on error messages from the literature that had been found to be confusing to students. On the other hand, we acknowledge that newer versions of Python have included improvements to some of the error messages we analyzed. For example, some of the code snippets we used that resulted in an ``invalid token'' error would have resulted into a ``SyntaxError: leading zeros in decimal integer literals are not permitted; use an 0o prefix for octal integers'' with newer Python versions. We consider the latter to be easier to understand for novice programmers.

Regarding the code snippets used in the analysis, they were created by the authors and were not student code. It is possible that the performance of Codex in explaining error messages for student code would be different. In our future work, we are interested in studying the error message explanations with student programs and with student evaluators. In addition, most of the source codes were relatively short. The performance of large language models in explaining error messages might be affected by the length or the complexity of source code, which future work should examine in greater detail. Similarly, our code snippets only included singular errors -- future work could analyze how well large language models can explain error messages when the source code that leads to those messages contains multiple issues.

When prompting Codex to generate an explanation of the error message and a fix to the program, we asked for both the explanation and the fix with a single prompt (``Plain English explanation of why does running the above code cause an error and how to fix the problem''). Performance could have increased had we asked for these separately. In addition, we did not give any examples of good error message explanations and fixes to the code in the prompt -- i.e. we relied on ``zero shot learning''~\cite{liu2021pre}. Prior work has found that giving even just a few examples (i.e. ``few shot learning'') can drastically improve the performance of large language models~\cite{brown2020language}.

\section{Conclusion}
\label{sec:conclusion}

We used large language models to try improve programming error messages (PEMs). We collected Python error messages that had been reported as most unreadable in prior work~\cite{denny2021designing,becker2021towards} and generated code examples that produced these error messages. We conducted prompt engineering using OpenAI Codex to identify prompts that would produce explanations of the PEMs and actionable fixes that could be applied to the code examples to fix the error. We evaluated the explanations and fixes created to examine whether they have utility in introductory programming classrooms. To summarize, we answer our research questions as follows.

\textit{RQ1: How well can Codex explain different error messages?} 
Overall, the explanations created by Codex were quite comprehensible (88\%). Codex produced an output with an explanation to 84\% of the provided codes and error messages, but only about half (57\%) of these explanations were deemed correct (48\% of all inputs). 

\textit{RQ2: What is the quality of the code fix suggestions that Codex generates?} 
Although 70\% of the outputs had a proposed fix, a little less than half (47\%) of those were deemed correct (33\% of all inputs). 

While the above results are aggregated over different PEMs, program categories, and Codex temperature values, we found cases where Codex seems to perform better. For example, we noticed that the results were better across the board when using the temperature value of 0. Similarly, we found that there were certain cases where Codex was more likely to provide faulty explanations and suggest fixes that are incorrect, and highlighted a potential way around this by having a two-step system that would look into the error message and the complexity of the source code before deciding whether to use LLMs or other more traditional support mechanisms.

The key implications of this work are that programming error message explanations and suggested fixes generated by LLMs are not yet ready for production use in introductory programming classes, as there are risks that students may interpret potentially faulty LLM outputs as coming from an authority, and end up attempting to fix their programs in ways that do not actually help. At the same time, our results show that LLMs could be a useful tool for improving PEMs, although additional effort needs to be taken both when using LLMs to enhance the error messages and when coming up with ways to produce high-quality enhancements. Enhancing programming error messages could help students in debugging their programs as traditional error messages are often cryptic and hard to understand for novice programmers~\cite{denny2021designing,denny2020error}.

The present results were obtained with the \texttt{code-davinci-002} model of OpenAI Codex, which was the most recent and performant Codex model at the time of the study. As LLMs improve over time, these results create a baseline that future model performance can be compared to. Future work should look in more depth into prompt engineering, for example by considering including the problem statement and perhaps a sample solution into the input, as well as look into applying and evaluating the enhanced programming error messages in classroom settings.


\bibliographystyle{ACM-Reference-Format}
\bibliography{references}
%\balance

\end{document}